\documentclass[10pt,twocolumn,letterpaper]{article}

\usepackage{cvpr}
\usepackage{times}
\usepackage{epsfig}
\usepackage{graphicx}
\usepackage{float}
\usepackage{amsmath}
\usepackage{amssymb}
\usepackage{float}
\usepackage{stfloats}

\usepackage[breaklinks=true,bookmarks=false]{hyperref}
\usepackage{amsmath}

\cvprfinalcopy 

\def\cvprPaperID{4} 

\begin{document}

\title{Human Perceptual Evaluations for Image Compression}

\author{Yash Patel\textsuperscript{2}, Srikar Appalaraju\textsuperscript{1}, R. Manmatha\textsuperscript{1}\\
	\textsuperscript{1}Amazon\\
	\textsuperscript{2}Center for Machine Perception, Czech Technical University, Prague, Czech Republic\\
	{\tt\small patelyas@cmp.felk.cvut.cz, (srikara,manmatha)@amazon.com}
}

\maketitle

\begin{abstract}
  Recently, there has been much interest in deep learning techniques to do image compression and there have been claims that several of these produce better results than engineered compression schemes (such as JPEG, JPEG2000 or BPG). A standard way of comparing image compression schemes today is to use perceptual similarity metrics such as PSNR or MS-SSIM (multi-scale structural similarity). This has led to some deep learning techniques which directly optimize for MS-SSIM by choosing it as a loss function. While this leads to a higher MS-SSIM for such techniques, we demonstrate using user studies that the resulting improvement may be misleading. Deep learning techniques for image compression with a higher MS-SSIM may actually be perceptually worse than engineered compression schemes with a lower MS-SSIM. 

\end{abstract}



\section{Introduction}
\label{sec:introduction}

Images have a lot of redundancy. Compression takes advantage of this redundancy to reduce image sizes. The value of compression is that it reduces the storage space required for images and the network bandwidth for transmitting them. It can also be used to reduce network latency when transmitting images. This can be significant since large numbers of photographs are transmitted and stored every day. For example, it is reported that users upload 300 million photographs to Facebook every day and Snapchat users share about 527,000 images per minute \cite{forbes}. Compression may be lossless (e.g. formats such as PNG \cite{boutell1997png}) or lossy (e.g. JPEG \cite{wallace1992jpeg}, JPEG2000  \cite{skodras2001jpeg} or BPG \cite{bpg}). Lossless compression implies that the original image can be perfectly reconstructed. Lossy compression on the other hand trades-off reconstruction error for improved compression (reduced file sizes). Traditionally, compression schemes have been engineered.  For example formats such as JPEG \cite{wallace1992jpeg}, JPEG2000 \cite{skodras2001jpeg} or BPG \cite{bpg} do not learn from the data. 

More recently, deep learned compression schemes \cite{balle2016end,balle2018variational,mentzer2018conditional,rippel2017real,lee2018context} have been proposed which can be trained using  data. The typical deep learned compression scheme consists of an encoder-decoder architecture with a loss function. The loss function usually optimizes for a distortion loss while at the same time trying to minimize the bit rate aka rate-distortion trade-off \cite{shannon1948mathematical}. The encoder maps the image into an embedding which is a much more compact representation. An entropy coder takes this embedding as input, quantizes it and converts it to a bitstream which is much more compact. To get back the image, entropy coding is reversed to produce an embedding which is then fed into a decoder to give a reconstructed approximate image as output (this is lossy compression). A variety of deep learning techniques have been proposed using the above basic framework.

Compression techniques are evaluated using measures such as structural similarity (SSIM) or PSNR between the original and reconstructed images. More recently, the trend is to evaluate compression using multiscale structure similarity (MS-SSIM) \cite{wang2003multiscale} sample compression works \cite{balle2016end,balle2018variational,mentzer2018conditional,rippel2017real,lee2018context}. PSNR and MS-SSIM were originally formulated as perceptual metrics but don't seem to completely capture certain type of distortions created by learned compression methods.

\begin{figure*}
	\centering
	\includegraphics[width=\textwidth]{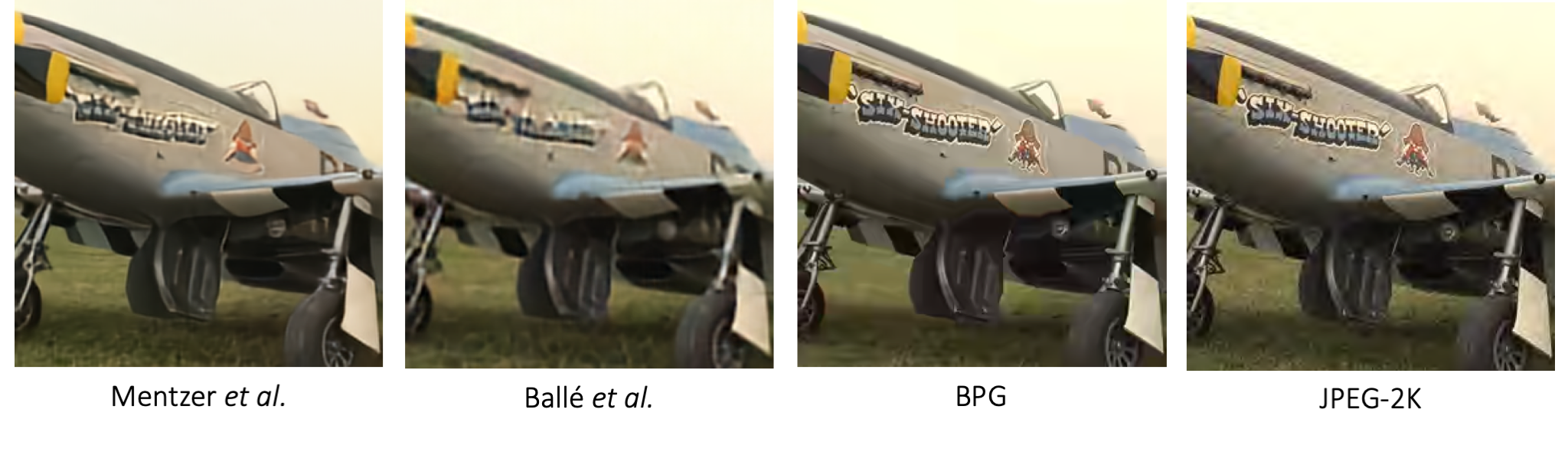}
	\caption{An example from the Kodak dataset. In order of MS-SSIM values Mentzer ~\etal $>$ Ballé ~\etal $>$ BPG $>$ JPEG-2000. However, visually the foreground and text from BPG and JPEG-2000 are clearly better in quality. The text from the deep learning techniques is hard to read.}
	\label{fig:kodak_samples}
\end{figure*}

While originally formulated as a perceptual loss metric, a  common trend has been to use MS-SSIM directly as a loss function for deep learning techniques and optimize for this measure \cite{mentzer2018conditional,rippel2017real,johnston2017improved}. Although, some methods \cite{balle2018variational,lee2018context} train separate models using MS-SSIM and MSE (mean-squared-error) loss functions. While the models trained for MSE get higher PSNR (compared to ones trained for MS-SSIM), they have lower MS-SSIM scores. A different ranking of methods may be produced by directly optimizing on MS-SSIM.  Several claims have been made that such approaches are better compared to engineered compression formats due to their higher MS-SSIM. We show that when we look at human perceptual judgments, PSNR and MS-SSIM scores may be misleading and may lead to the wrong conclusions about which technique is better. MS-SSIM cannot differentiate between a (locally) blurry patch and a patch which isn't blurry so this could be one reason for this difference. However, it is clear by looking at the images and user judgments that the problems with these metrics go beyond this issue.

In this paper we conduct extensive human study on perceptual similarity for  image compression techniques. Using MTurk platform, we make pairwise comparisons for the following methods: JPEG-2000  \cite{skodras2001jpeg}, BPG \cite{bpg}, Mentzer ~\etal \cite{mentzer2018conditional} and Ballé ~\etal \cite{balle2016end}. We demonstrate that learning based compression methods \cite{mentzer2018conditional,balle2016end} despite having higher MS-SSIM scores are visually worse, see Figure \ref{fig:kodak_samples} for an example. We pick \cite{mentzer2018conditional,balle2016end} as two methods for evaluation since their implementations are publicly available unlike a number of other methods. Refer to the original papers Mentzer ~\etal \cite{mentzer2018conditional} and Ballé ~\etal \cite{balle2016end} for detailed method descriptions.


\section{Human Evaluation Setup}
\label{sec:human_eval}

\begin{figure}
	\centering
	\includegraphics[width=0.45\textwidth]{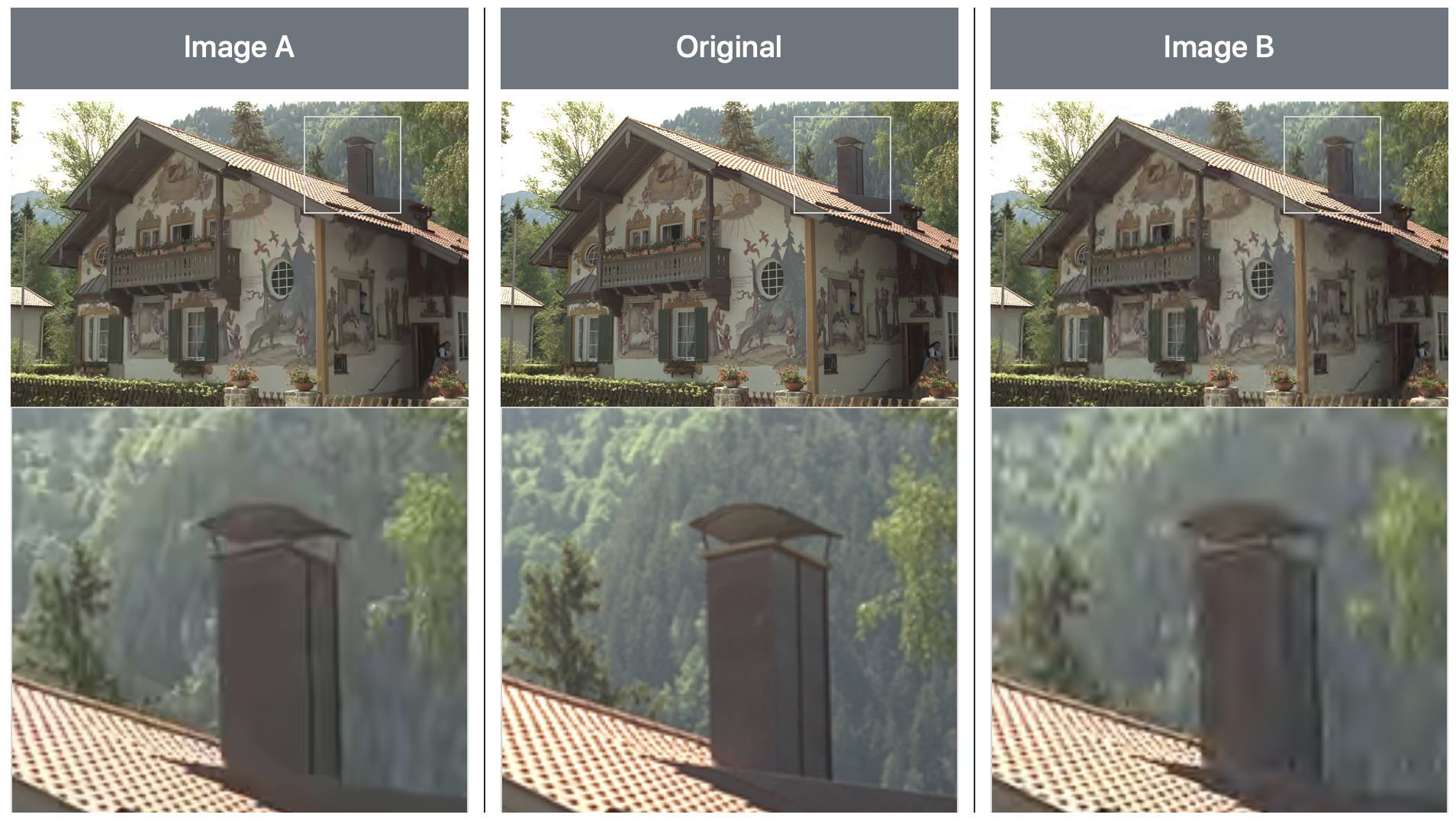}
	\caption{Sample instance from MTurk HIT. Entire images are shown at the top with original image in the middle and image from one method on the left and other on the right. The bottom images are magnified version of a small window which can be controlled by moving the cursor.}
	\label{fig:hit_example}
\end{figure}

We conduct human perceptual similarity study on Amazon MTurk by showing an evaluator original image along with two reconstructed images from two compression techniques at a time. They were asked to choose an image which is more similar to the original. The evaluators are shown entire images along with a synchronized (on all three) magnifying glass to observe finer details. This gives them a global context of the whole image and at the same time provides a quick way to access local regions. No time limit was placed for this human experiment. An instance from this evaluation page is shown in Figure \ref{fig:hit_example}.

The evaluators are forced to choose the compressed image which is closer to the original (2AFC). 2AFC is a known way of performing perceptual similarity evaluation and has been used by \cite{sajjadi2017enhancenet} for evaluating super-resolution techniques. In this setup, we compare two engineered (JPEG-2000 \cite{skodras2001jpeg}, BPG \cite{bpg}) and two learning based (Mentzer ~\etal \cite{mentzer2018conditional}, Ballé ~\etal \cite{balle2016end}) compression techniques by choosing all possible combinations (six pairs in total). Further, we do this at four different compression levels - i.e. bits-per-pixel (bpp) values: $0.23, 0.37, 0.67, 1.0$. The study is conducted on four standard datasets: Kodak \cite{kodak}, Urban100 \cite{huang2015single}, Set14 \cite{zeyde2010single} and Set5 \cite{bevilacqua2012low}.

With this setup, we have a total of $3432$ pairs ($6$ pairs for four methods, $4$ bpp values and $143$ images in total). For each such pair, we obtain $5$ evaluations resulting in a total of $17160$ HITs.

\begin{figure*}[h!]
	\centering
	\includegraphics[width=0.85\textwidth]{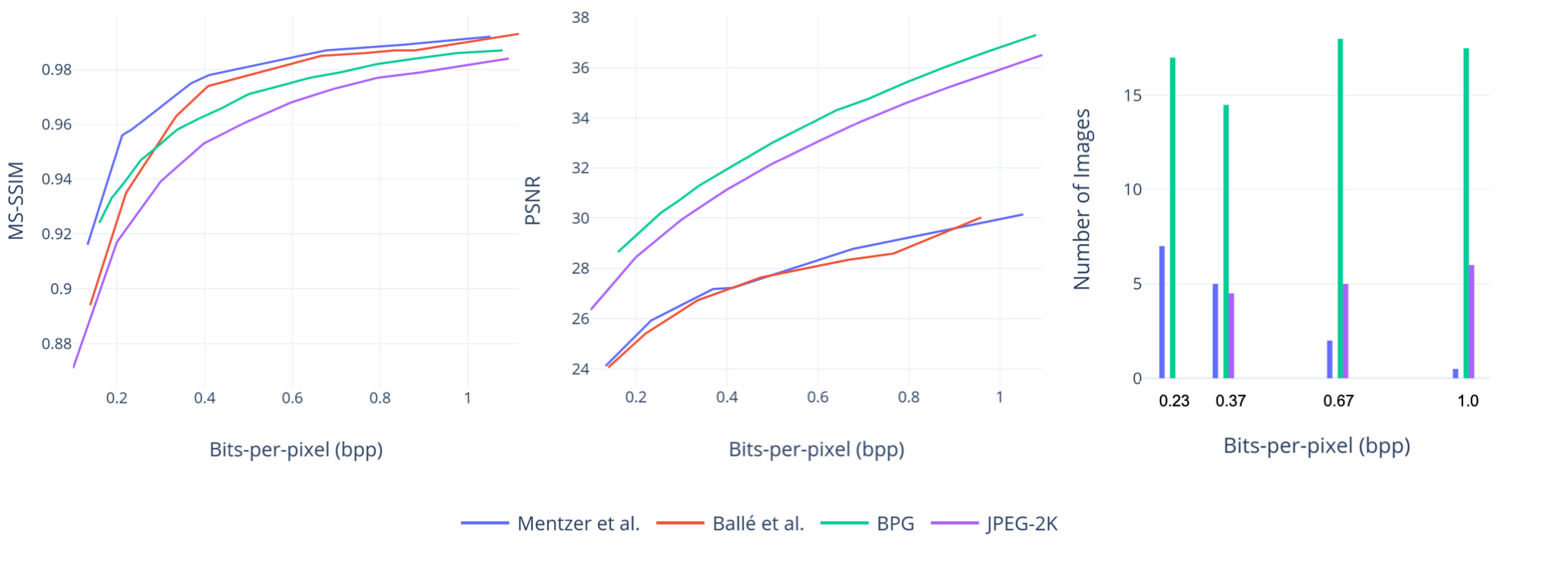}
	\caption{Evaluation on the Kodak dataset \cite{kodak} using MS-SSIM (left), PSNR (middle) and human study (right). In this case, Ballé ~\etal is never the best method for any image.}
	\label{fig:kodak_plots}
\end{figure*}

\begin{figure*}[h!]
	\centering
	\includegraphics[width=\textwidth]{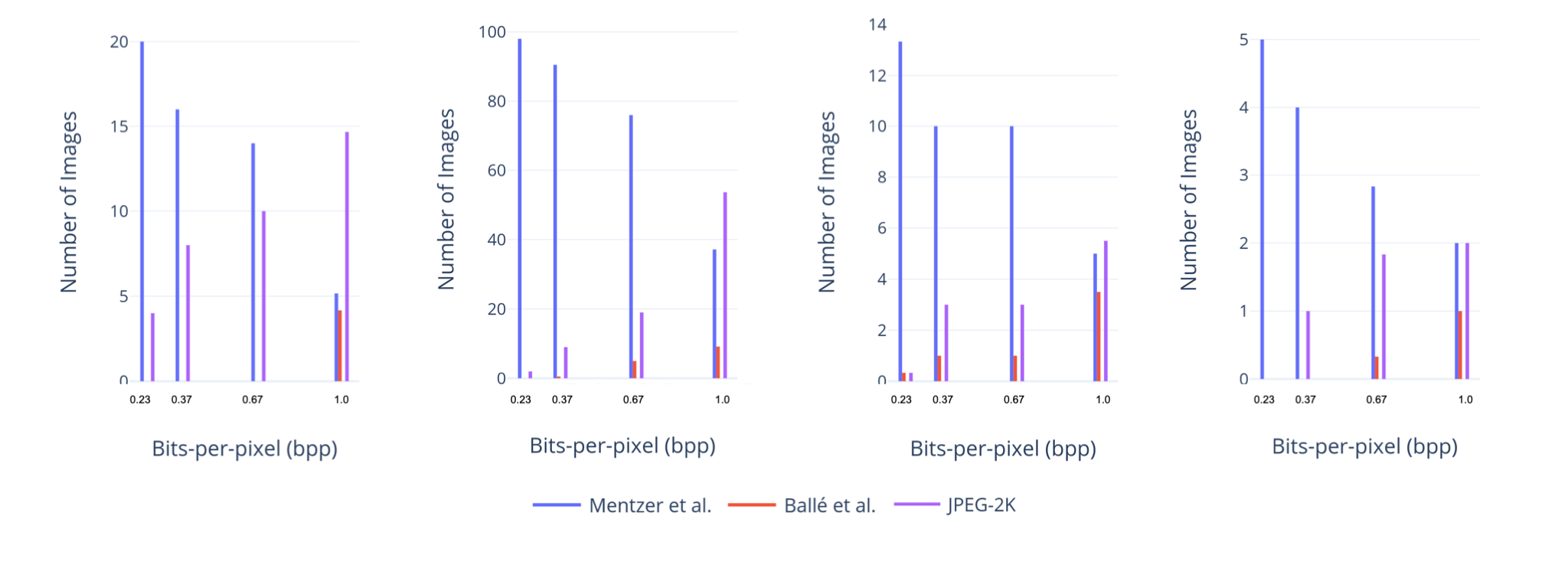}
	\caption{Human study on the Kodak (left most) \cite{kodak}, Urban100 (second from left) \cite{huang2015single}, Set14 (second from right) \cite{zeyde2010single} and Set5 (right most) \cite{bevilacqua2012low} when BPG is excluded from the comparison.}
	\label{fig:human_eval_rm_bpg}
\end{figure*}

\begin{figure*}[h!]
	\centering
	\includegraphics[width=0.85\textwidth]{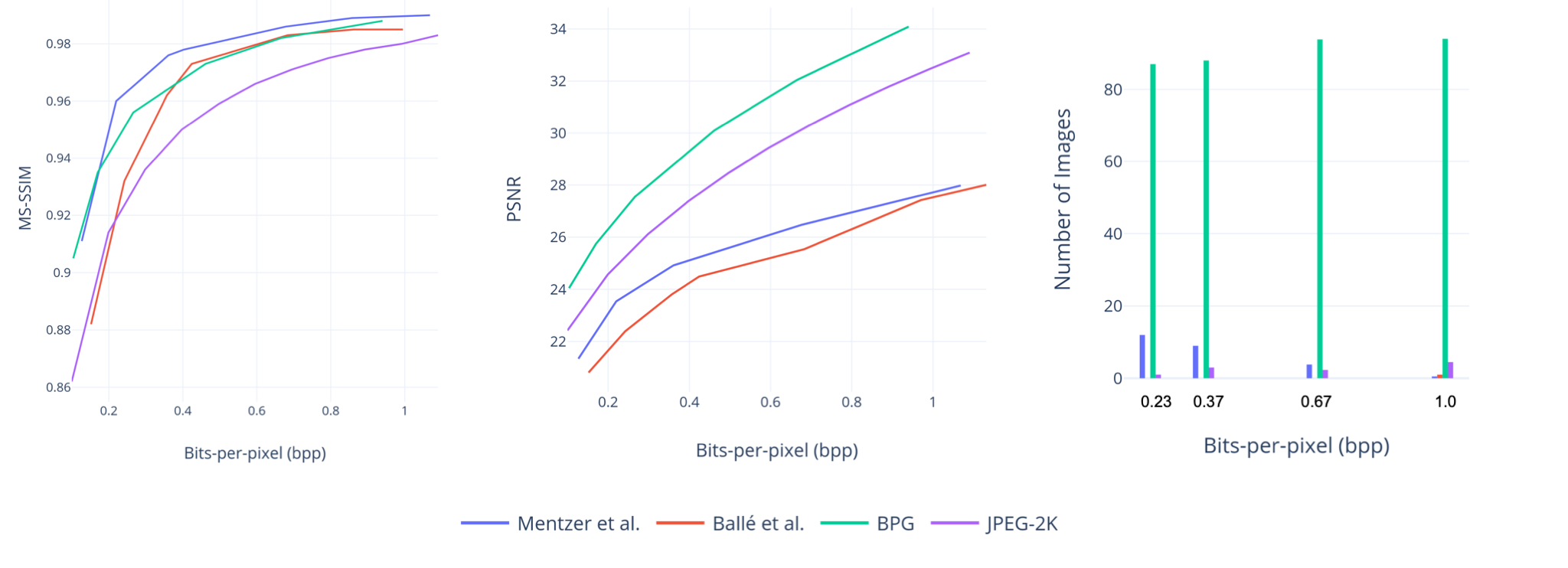}
	\caption{Evaluation on the Urban100 dataset \cite{huang2015single} using MS-SSIM (left), PSNR (middle) and human study (right).}
	\label{fig:urban100_plots}
\end{figure*}

\begin{figure*}[h!]
	\centering
	\includegraphics[width=0.85\textwidth]{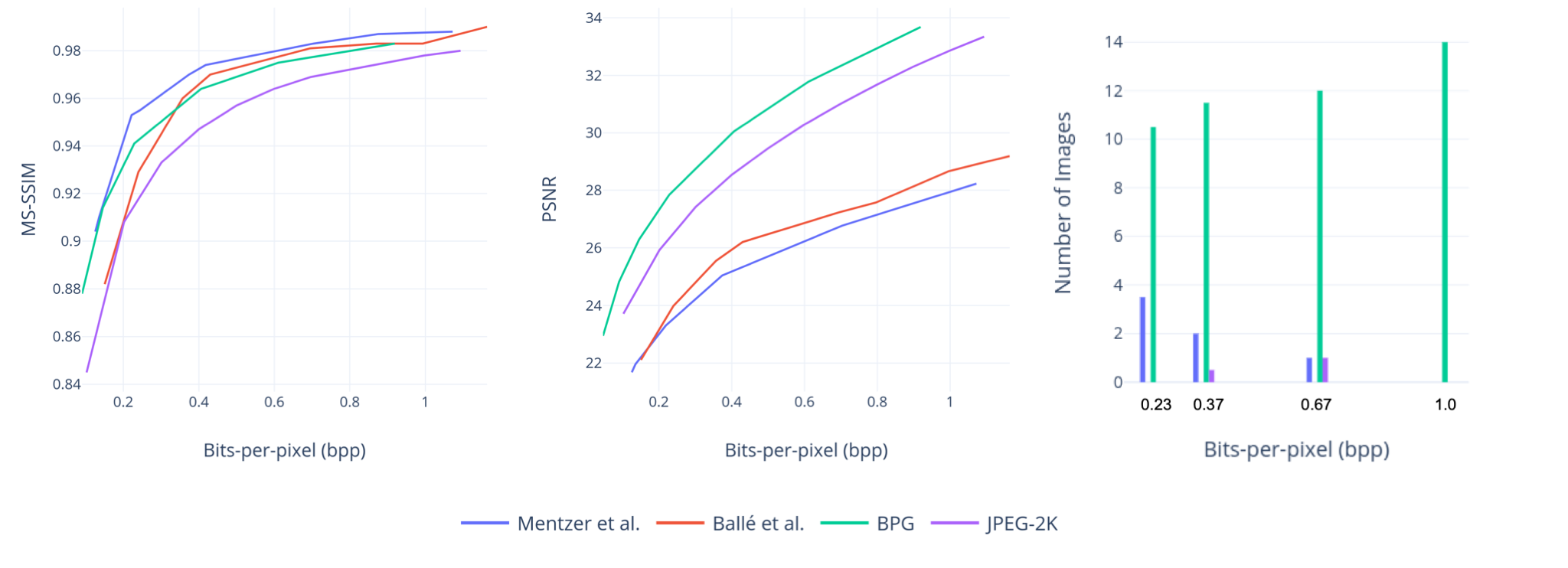}
	\caption{Evaluation on Set14 dataset \cite{zeyde2010single} using MS-SSIM (left), PSNR (middle) and human study (right). In this case, Ballé ~\etal is never the best method for any image.}
	\label{fig:set14_plots}
\end{figure*}

\begin{figure*}[h!]
	\centering
	\includegraphics[width=0.85\textwidth]{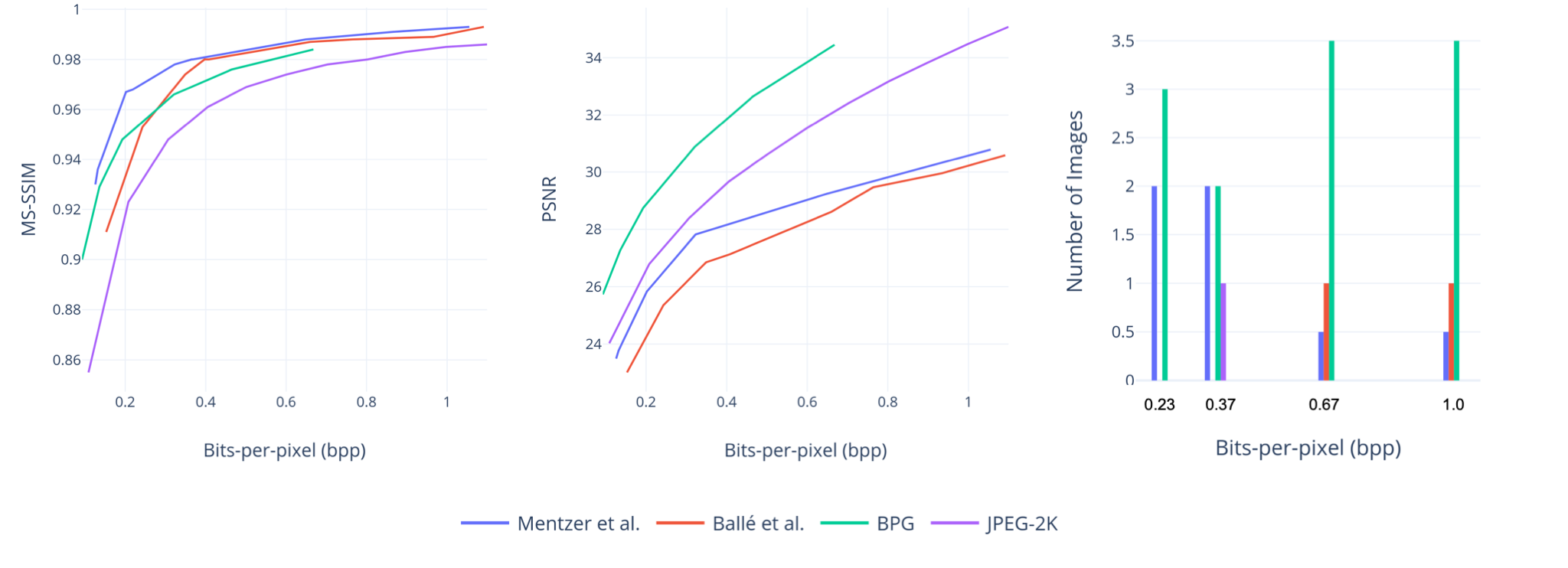}
	\caption{Evaluation on Set5 dataset \cite{bevilacqua2012low} using MS-SSIM (left), PSNR (middle) and human study (right).}
	\label{fig:set5_plots}
\end{figure*}

\section{Results}
\label{sec:results}
By varying the hyper-parameters such as the number of channels in the  bottleneck, weight for distortion loss (target bit-rate for \cite{mentzer2018conditional}), we reproduce the models for \cite{mentzer2018conditional,balle2016end} at different bpp's by training on the ImageNet dataset \cite{russakovsky2015imagenet}. For each test dataset, we compress all the images using a model and take an average of bpp, MS-SSIM and PSNR across the images within the dataset. We do the same for all the models which gives us multiple points on the MS-SSIM vs bpp and PSNR vs bpp curves. We interpolate the values between two such points and we do not extrapolate the values outside the bpp range \footnote{in the case of \cite{balle2016end} we used an MS-SSIM loss instead of the MSE loss in the original paper but this does not change the general conclusions of the paper}.

For human evaluation, based on the setup explained in Section \ref{sec:human_eval}, for each image and a given bpp value we have $5$ pair-wise votes in the form of method-A vs method-B. Since we have all possible pairs for four methods under consideration, we aggregate these votes and obtain the method which does best for the given image (at a particular bpp value) based on maximum votes. In a rare case of a tie between two or more methods (happens for less than $2\%$ of cases), we assign a score of $0.5$ (two-method tie) or $0.33$ (three-method tie) to each tying methods. In the Figures \ref{fig:kodak_plots}, \ref{fig:urban100_plots}, \ref{fig:set14_plots} and \ref{fig:set5_plots}, we show the number of images (y-axis) for which a method performs best.

The comparisons for Kodak \cite{kodak} are shown in Figure \ref{fig:kodak_plots}. We observe that Mentzer ~\etal \cite{mentzer2018conditional} have the highest MS-SSIM score for all bpp values. Ballé ~\etal \cite{balle2016end} obtains higher MS-SSIM score compared to BPG and JPEG-2000 after $0.3$ bpp, which is followed by BPG and JPEG-2000 has the lowest MS-SSIM scores. Contrary to the MS-SSIM scores, human evaluation shows that BPG performs significantly better than all the other methods, including \cite{mentzer2018conditional} which has substantially higher MS-SSIM. In Figure \ref{fig:human_eval_rm_bpg} we show the human study plots when BPG is excluded from the comparison. In this case, Mentzer ~\etal \cite{mentzer2018conditional} does best at lower bpp and JPEG-2000 does best at higher bpp. This observation contradicts the PSNR scores which are higher for JPEG-2000. Thereby for learned approaches, MS-SIM is still a better loss function than PSNR.

Similarly comparisons for other datasets are made in: Figure. \ref{fig:urban100_plots} for Urban100 \cite{huang2015single}, \ref{fig:set14_plots} for Set14 \cite{zeyde2010single} and Figure \ref{fig:set5_plots} for Set5 \cite{bevilacqua2012low}.


\section{Discussion}

Deep learning techniques directly optimize metrics such as MS-SSIM or MSE (mean-squared-error) and hence are able to get high numbers on these metrics. The experiments above show that there is no direct correlation between the quality of output from a given compression technique and metrics such as PSNR or MS-SSIM. While we picked two techniques this seems to be a more general problem for deep learned image compression techniques. In the absence of a better metric, papers should provide an implementation or at least images on standard datasets so that researchers may compare their new technique against an existing one. Having a better perceptual metric close to human visual perception would be ideal. Note that similar observations have been drawn in \cite{2019arXiv190708310P}.

\section*{Acknowledgment}
We would like to thank Joel Chan and Peter Hallinan for helping us in setting up the human evaluations.

{\small
\bibliographystyle{ieee}
\bibliography{example_paper}

\begin{thebibliography}{10}\itemsep=-1pt

\bibitem{balle2016end}
J.~Ball{\'e}, V.~Laparra, and E.~P. Simoncelli.
\newblock End-to-end optimized image compression.
\newblock {\em arXiv preprint arXiv:1611.01704}, 2016.

\bibitem{balle2018variational}
J.~Ball{\'e}, D.~Minnen, S.~Singh, S.~J. Hwang, and N.~Johnston.
\newblock Variational image compression with a scale hyperprior.
\newblock {\em arXiv preprint arXiv:1802.01436}, 2018.

\bibitem{bpg}
F.~Bellard.
\newblock Bpg image format, 2014.

\bibitem{bevilacqua2012low}
M.~Bevilacqua, A.~Roumy, C.~Guillemot, and M.~L. Alberi-Morel.
\newblock Low-complexity single-image super-resolution based on nonnegative
  neighbor embedding.
\newblock 2012.

\bibitem{boutell1997png}
T.~Boutell.
\newblock Png (portable network graphics) specification version 1.0.
\newblock Technical report, 1997.

\bibitem{forbes}
Forbes.
\newblock How much data do we create every day? the mind-blowing stats everyone
  should read, 2018.

\bibitem{kodak}
R.~Franzen.
\newblock Kodak lossless true color image suite.
\newblock {\em source: http://r0k. us/graphics/kodak}, 4, 1999.

\bibitem{huang2015single}
J.-B. Huang, A.~Singh, and N.~Ahuja.
\newblock Single image super-resolution from transformed self-exemplars.
\newblock In {\em Proceedings of the IEEE Conference on Computer Vision and
  Pattern Recognition}, 2015.

\bibitem{johnston2017improved}
N.~Johnston, D.~Vincent, D.~Minnen, M.~Covell, S.~Singh, T.~Chinen, S.~J.
  Hwang, J.~Shor, and G.~Toderici.
\newblock Improved lossy image compression with priming and spatially adaptive
  bit rates for recurrent networks.
\newblock {\em structure}, 10:23, 2017.

\bibitem{lee2018context}
J.~Lee, S.~Cho, and S.-K. Beack.
\newblock Context-adaptive entropy model for end-to-end optimized image
  compression.
\newblock {\em ICLR}, 2019.

\bibitem{mentzer2018conditional}
F.~Mentzer, E.~Agustsson, M.~Tschannen, R.~Timofte, and L.~Van~Gool.
\newblock Conditional probability models for deep image compression.
\newblock In {\em Proceedings of the IEEE Conference on Computer Vision and
  Pattern Recognition}, pages 4394--4402, 2018.

\bibitem{2019arXiv190708310P}
Y.~{Patel}, S.~{Appalaraju}, and R.~{Manmatha}.
\newblock {Deep Perceptual Compression}.
\newblock {\em arXiv preprint arXiv:1907.08310}.

\bibitem{rippel2017real}
O.~Rippel and L.~Bourdev.
\newblock Real-time adaptive image compression.
\newblock {\em arXiv preprint arXiv:1705.05823}, 2017.

\bibitem{russakovsky2015imagenet}
O.~Russakovsky, J.~Deng, H.~Su, J.~Krause, S.~Satheesh, S.~Ma, Z.~Huang,
  A.~Karpathy, A.~Khosla, M.~Bernstein, et~al.
\newblock Imagenet large scale visual recognition challenge.
\newblock {\em International journal of computer vision}, 2015.

\bibitem{sajjadi2017enhancenet}
M.~S. Sajjadi, B.~Sch{\"o}lkopf, and M.~Hirsch.
\newblock Enhancenet: Single image super-resolution through automated texture
  synthesis.
\newblock In {\em Computer Vision (ICCV), 2017 IEEE International Conference
  on}, pages 4501--4510. IEEE, 2017.

\bibitem{shannon1948mathematical}
C.~E. Shannon.
\newblock A mathematical theory of communication.
\newblock {\em Bell system technical journal}, 27(3):379--423, 1948.

\bibitem{skodras2001jpeg}
A.~Skodras, C.~Christopoulos, and T.~Ebrahimi.
\newblock The jpeg 2000 still image compression standard.
\newblock {\em IEEE Signal processing magazine}, 2001.

\bibitem{wallace1992jpeg}
G.~K. Wallace.
\newblock The jpeg still picture compression standard.
\newblock {\em IEEE transactions on consumer electronics}, 1992.

\bibitem{wang2003multiscale}
Z.~Wang, E.~P. Simoncelli, and A.~C. Bovik.
\newblock Multiscale structural similarity for image quality assessment.
\newblock In {\em The Thrity-Seventh Asilomar Conference on Signals, Systems \&
  Computers, 2003}, volume~2, pages 1398--1402. Ieee, 2003.

\bibitem{zeyde2010single}
R.~Zeyde, M.~Elad, and M.~Protter.
\newblock On single image scale-up using sparse-representations.
\newblock In {\em International conference on curves and surfaces}. Springer,
  2010.

\end{thebibliography}
}

\end{document}